\documentclass[pre,twocolumn,showpacs]{revtex4}
\usepackage{amsmath}

\usepackage{color}

\begin{document}

\title{A self-propelled particle in an external potential: is there an effective 
temperature?}

\author{Grzegorz Szamel}
\affiliation{Department of Chemistry, 
Colorado State University, Fort Collins, CO 80525}

\date{\today}


\begin{abstract}
We study a stationary state of a single self-propelled, athermal particle in linear
and quadratic external potentials. The self-propulsion is modeled as 
a fluctuating force evolving according to the Ornstein-Uhlenbeck process, 
independently of the state of the particle. Without an external potential, 
in the long time limit, the self-propelled particle moving in a viscous 
medium performs diffusive motion, which allows one to identify 
an effective temperature. We show that 
in the presence of a linear external potential the stationary state distribution
has an exponential form with the sedimentation length 
determined by the effective temperature of the free self-propelled particle.
In the presence of a quadratic external potential the stationary
state distribution has a Gaussian form. However, in 
general, this distribution is not determined by the 
effective temperature of the free self-propelled particle.
\end{abstract} 

\maketitle

\section{Introduction}\label{sec:intro}

Recently, there has been a lot of interest in the static and dynamic properties 
of particles that are self-propelled and, thus, can move on their own 
accord \cite{Ramaswamyrev,Romanczukrev,Catesrev}. 
These particles are said to move \emph{actively} and to form \emph{active matter}.
      
There are two motivations for the interest in active matter systems. First,
these systems model static and dynamic properties of specific biological and physical
systems in which self-propelled motion occurs. For example, a system of particles
with the so-called run-and-tumble motion serves as a model for 
\textit{Escherichia coli} bacteria \cite{Catesrev}. 
Similarly, a system of so-called active Brownian
particles \cite{Romanczukrev} is a model system for Janus colloidal particles
\cite{Erbe}.  
The second motivation for the interest in active matter systems is the fundamental
fascination with non-equilibrium physical systems and, in particular, with 
systems without detailed balance. 

The present contribution is inspired by recent studies that showed that, at least
in some cases, active matter systems can exhibit phenomena that are commonly
found in standard (thermal, non-active) systems. For example, Palacci \textit{et al.}
\cite{Palacci} found that a dilute active colloidal suspension under gravity
exhibits qualitatively the same exponential density distribution as a standard 
dilute thermal colloidal system. Notably, the parameter that replaces the thermal
system's temperature coincides with the effective temperature that was inferred
from an independent measurement of the long-time diffusive motion of an active
colloidal particle. More interestingly, behavior similar to that common in thermal 
systems was found in systems consisting of interacting active particles. 
For example, Bialk\'{e} \textit{et al.} \cite{Bialke1} used computer simulations to 
show that a system of active Brownian particles can crystallize at sufficiently 
high densities. Next, Das \textit{et al.} \cite{Das} used both a computer
simulation and an integral equation theory to show that activity promotes
phase separation in an active binary mixture. Finally, it was found that 
active systems can exhibit glassy dynamics. Berthier and Kurchan
\cite{BerthierKurchan} analyzed a simple model active system inspired by the so-called
spherical $p$-spin model and showed that it can exhibit kinetic arrest. This
pioneering study was followed by two computer simulation investigations of
systems of active Brownian particles \cite{Ni,Berthier} which showed that, 
generically, active systems exhibit glassy dynamics, but 
the onset of glassy behavior is pushed towards higher densities compared
with systems of non-active particles. In turn, the latter simulations 
inspired a very recent mode-coupling-like description of glassy dynamics in
active systems \cite{Farage}. 

Results of some of the investigations mentioned above \cite{Palacci,Das}  
suggest an emergence of effective thermal behavior and, more importantly,
effective temperature \cite{Loi}. 
It should be noted, however, that other studies \cite{Bialke1,Fily} question
the usefulness of the notion of effective temperature. In particular, 
Fily and Marchetti \cite{Fily} argue that this notion holds only in the dilute limit. 

Our goal is to test the validity of effective temperature in a simple model. 
To keep the model exactly solvable we replace a system of interacting particles
by a single particle in an external field. 
Specifically, we compare the behavior of a single active particle without any external 
potential (for which an effective temperature can be easily defined) 
with the behavior of the same particle in two different external potentials.

There is a number of different models of self-propelled motion \cite{Romanczukrev}.
Their common feature is that an active particle moves under an influence of 
an internal self-propulsion force which evolves in some specified way, 
independently of the state of the particle. 
Here we will consider the continuous time, one-dimensional version of the model
introduced by Berthier \cite{Berthier,Levis}. 
In the original model of Ref. \cite{Berthier}
Monte-Carlo dynamics with correlated trial moves was used
(in standard Monte-Carlo dynamics subsequent trial moves are 
un-correlated \cite{AT}). In our model, the particle is subjected to a self-propulsion
force and, possibly, a conservative force originating from an external potential. 
The self-propulsion force has a vanishing average, a finite mean-square and a finite 
persistence (\textit{i.e.} relaxation) time. We choose a rather simple 
evolution for the self-propulsion: 
we assume that the self-propulsion force evolves according to the Ornstein-Uhlenbeck
stochastic process, independently of the state of the particle and, in particular, of 
any external force acting on it. Our choice of the self-propulsion force evolution
leads to relatively simple equations of motion for the probability 
distribution of the active particle, with 
stationary state distributions that can be derived analytically.

We will be mostly concerned with the stationary state distributions under the
influence of an external potential. In particular, we will show that even though
the self-propulsion force evolves on its own, non-trivial correlations between
the position of the particle and the self-propulsion force can develop. 

We start by briefly discussing the motion of a free self-propelled
particle. We note that by using the theoretical apparatus developed to
analyze Brownian motion we can easily derive the long-time-scale 
description of the active motion and define the free particle effective temperature.
Next, we analyze the self-propelled particle under the influence of a constant
force. We show that the stationary state probability distribution
has the familiar exponential form and that it can be expressed in terms of the 
free particle's effective temperature. 
Finally, we analyze the self-propelled 
particle under the influence of a harmonic force. We show that in this case 
the stationary state probability distribution has the familiar Gaussian form.
However, in general, it is not determined by the effective temperature obtained
from the motion of the free self-propelled particle. We also examine 
an effective temperature defined through a fluctuation-dissipation relation. 
We end the paper with a brief discussion of the
results, which should be applicable in a broader context.

\section{Free self-propelled particle}\label{sec:free}

The free self-propelled particle moves in a viscous medium under the influence 
of an internal self-propulsion force. We assume that viscous dissipation dominates 
and consequently the motion is over-damped. The medium is characterized by the 
single-particle friction coefficient $\xi_0$. We assume that any random force
originating from the solvent's fluctuations is negligible and, thus, the particle 
is non-Brownian. The self-propulsion force evolves
according to the Ornstein-Uhlenbeck stochastic process. Specifically, the 
force relaxes to its average zero value on the time scale characterized by the 
inverse rate $\gamma^{-1}$ and it also changes by random, uncorrelated increments
due to an internal noise. As a consequence, the self-propulsion
force acquires a finite, non-zero mean-square. 

The time evolution of the system is described by the following equations of motion,
\begin{eqnarray}\label{freepos}
\partial_t x(t) &=& \xi_0^{-1} f(t),
\\ \label{force}
\partial_t f(t) &=& -\gamma f(t) + \eta(t).
\end{eqnarray}
Eq. (\ref{freepos}) describes over-damped motion of the particle and Eq. (\ref{force})
describes the evolution of the self-propulsion force. In Eq. (\ref{force}) 
$\eta(t)$ is a white Gaussian noise with the auto-correlation function given by
\begin{eqnarray}
\left<\eta(t) \eta(t')\right>_{noise} = 2 D_f \delta(t-t'), 
\end{eqnarray} where $\left< \dots \right>_{noise}$ denotes averaging of 
a Gaussian white noise $\eta$. 

Equivalently, the motion of the self-propelled 
particle can be described by a joint probability distribution
for the particle's position and the self-propulsion force, $P(x,f;t)$.
The equation of motion for this distribution reads:
\begin{eqnarray}\label{probfree}
\partial_t P(x,f;t) &=& -\frac{f}{\xi_0}\frac{\partial P(x,f;t)}{\partial x}
\\ \nonumber && 
+\frac{\partial}{\partial f}\left(\gamma f P(x,f;t) + D_f \frac{\partial}{\partial f}
P(x,v;t)\right).
\end{eqnarray}

It can be easily showed that in the
stationary state 
\begin{equation}
P^{ss}(x,f)\propto \exp\left(-\frac{\gamma f^2}{2 D_f}\right)
\end{equation}
and $\left<f^2\right>=D_f/\gamma$. Here and in the following 
$\left< \dots \right>$ denotes 
averaging over the stationary distribution of the position and self-propulsion force.

We note that Eq. (\ref{probfree}) is equivalent to the so-called Fokker-Planck
equation that describes the motion of a Brownian particle on a time scale on which
its velocity relaxation can be observed \cite{VanKampen}. Indeed, replacing 
$f/\xi_0$ by particle's velocity $v$ changes Eq. (\ref{probfree}) into the
Fokker-Planck equation. Consequently, we can use the well known theoretical
analyzes of Brownian motion \cite{VanKampen,Titulaer} for the present case of
self-propelled motion. We see immediately that the long-time motion of the 
self-propelled particle is diffusive and the diffusion constant is equal to
\begin{equation}\label{difffree}
D=\left<f^2\right>/\left(\xi_0^2 \gamma\right)=D_f/(\xi_0\gamma)^2.
\end{equation}
Since the particle is moving in a viscous medium, using the standard Einstein 
relation between the temperature and friction, and diffusion coefficients allows
us to define an effective temperature,
\begin{equation}\label{Tefffree}
T_{eff} = D\xi_0 = D_f/\left(\xi_0\gamma^2\right)
\end{equation}
(we use a system of units in which the Boltzmann constant $k_B$ is equal to 1).

We should note that this long-time diffusive motion of the self-propelled
particle, with diffusion constant given by Eq. (\ref{difffree}), 
is established on the time scale much longer than $\gamma^{-1}$.
We might expect that the long-time motion may become different (or at least that
the diffusion constant becomes different from (\ref{difffree})) 
if there is another comparable
or shorter time scale in the problem. We should note in this context
that interesting self-propulsion-related phenomena are observed for 
slowly relaxing self-propulsion forces, \textit{i.e.} precisely when 
$\gamma^{-1}$ is \textit{not} the shortest time scale in the problem.

\section{Self-propelled particle under the influence
of a constant force: sedimentation}\label{sec:sed}

If there is an external, conservative, time-independent 
force acting on the particle, the equation of motion for 
the position of the self-propelled particle has the following form:
\begin{eqnarray}\label{potpos}
\partial_t x(t) &=& \xi_0^{-1} \left( f(t) +  F^{ext}(x(t))\right)
\end{eqnarray}
where $F^{ext}(x) = -\partial_x V^{ext}(x)$ is the 
external, conservative, time-independent force acting on the particle.
Eq. (\ref{potpos}) needs to be augmented by the equation of motion for the 
self-propulsion force, Eq. (\ref{force}). 
We emphasize that the evolution of the self-propulsion force is un-changed.

As in the case of a free self-propelled particle, 
we can describe the time dependence of the state of the particle through the joint
probability distribution of the position and the self-propulsion force,
\begin{eqnarray}\label{probpot}
\partial_t P(x,f;t) &=& -\frac{1}{\xi_0}\frac{\partial}{\partial x}\left(\left( 
f + F^{ext}(x)\right) P(x,f;t)\right)
\\ \nonumber && 
+\frac{\partial}{\partial f}\left(\gamma f P(x,f;t) + D_f \frac{\partial}{\partial f}
P(x,f;t)\right).
\end{eqnarray}
We should emphasize that since the self-propulsion force evolves independently 
of the external force, equation of motion (\ref{probpot}) is qualitatively 
\emph{different} from the
Fokker-Planck equation for the joint probability distribution of a position and
velocity of a Brownian particle moving under the influence of an external force. 
Thus, we cannot use the 
theoretical apparatus developed in Refs. \cite{VanKampen,Titulaer}.

In the remainder of this section we briefly analyze the stationary state
of a self-propelled particle under the influence of a constant 
external force, which models sedimentation in a dilute active colloidal suspension
\cite{Palacci}. In the next section we investigate a self-propelled 
particle in a harmonic potential.

For a single self-propelled particle under the influence of a constant gravitational
force, the equation of motion has the following form:
\begin{eqnarray}\label{probsed}
\partial_t P(x,f;t) &=& -\frac{1}{\xi_0}\frac{\partial}{\partial x}\left(\left( 
f - mg \right) P(x,f;t)\right)
\\ \nonumber && 
+\frac{\partial}{\partial f}\left(\gamma f P(x,f;t) + D_f \frac{\partial}{\partial f}
P(x,f;t)\right),
\end{eqnarray}
where $g$ is the gravitational acceleration and $m$ is the mass of the particle. 
Note that this equation is only valid above a lower wall, which we assume to
be located at $x=0$. In principle, Eq. (\ref{probsed}) has to be accompanied by a 
term that ensures that the current through the lower wall vanishes. This term
is not needed for finding the stationary state distribution since in the stationary
state the current vanishes everywhere. 

We note that the so-called drift coefficients \cite{VanKampen} in Eq. (\ref{probsed})
are linear in $x$ and $f$. This fact suggests looking for a stationary 
distribution having a Gaussian form. 
It can be showed that the following distribution is a stationary solution
of Eq. (\ref{probsed}):
\begin{equation}\label{ssprobsed}
P^{ss}(x,f) \propto \exp\left(-ax-bf^2-cf\right),
\end{equation}
where $a=mg\xi_0\gamma^2/D_f$,
$b=\gamma/(2D_f)$ and $c=-a/(\xi_0\gamma)$.

It follows from Eq. (\ref{ssprobsed}) that the stationary state distribution
of positions is exponential,
\begin{equation}\label{ssprobxsed}
P^{ss}(x) = \int df P^{ss}(x,f) \propto \exp(-x/\delta_{eff})
\end{equation}
where the so-called sedimentation length 
$\delta_{eff}=1/a=D_f/\left(mg\xi_0\gamma^2\right)$. We note that the
sedimentation length of a dilute system of non-active Brownian particles 
at temperature $T$ is given by $\delta=T/\left(mg\right)$. We can thus conclude that 
the sedimentation length of a dilute system of self-propelled particles has the
same form as that of non-active Brownian particles if instead of the 
equilibrium temperature one uses effective temperature 
$T_{eff}$ of a free self-propelled particle, $T_{eff}=D_f/\left(\xi_0\gamma^2\right)$.
This agrees with the experimental result of Palacci \textit{et al.} \cite{Palacci}
obtained for a slightly different system of the active Brownian
particles. We note that the consistency between the free particle effective
temperature and the sedimentation length is not obvious. In fact, 
Tailleur and Cates \cite{TC} showed that for the run-and-tumble model of active particles
the stationary state distribution in a linear potential has the exponential 
form, but the free particle effective temperature does not determine the 
sedimentation length.

According to Eq. (\ref{ssprobsed}), 
the self-propulsion distribution is different from that 
of a free self-propelled particle, 
\begin{eqnarray}
P^{ss}(f) &=& \int dx P^{ss}(x,f) 
\\ \nonumber &=& 
\left(\frac{b}{\pi}\right)^{1/2} \exp\left(-bf^2-cf - c^2/(4b)\right).
\end{eqnarray}
We note that to show that the above distribution
is consistent with Eq. (\ref{probsed}) 
one should pay attention to the $x=0$ boundary term.

In particular, in the present case there is a 
non-zero local stationary state self-propulsion:
\begin{equation}\label{lespsed}
\left<f\right>_{lss} = \int df f P^{ss}(f|x) = - \frac{c}{2b}=mg,
\end{equation}
where $\left< ... \right>_{lss}$ denotes the local stationary state average
or, more precisely, the stationary state average 
over self-propulsion under the condition 
that the particle is at position $x$. In other words, $P^{ss}(f|x)$ 
in Eq. (\ref{lespsed}) is the conditional stationary state distribution of 
the self-propulsion force,
\begin{equation}
P^{ss}(f|x) = P^{ss}(x,f)/P^{ss}(x),
\end{equation}
where $P^{ss}(x)$ is the stationary state distribution of particle's positions,
$P^{ss}(x)=\int df P^{ss}(x,f)$. 

Non-zero average 
self-propulsion follows from the condition that there
should be no net current in the stationary state. Let's define the current density,
\begin{equation}
\partial_t P(x;t) = -\partial_x j(x;t)
\end{equation}
where $P(x;t)=\int df P(x,f;t)$. Thus the current density is given by
\begin{equation}
\xi_0^{-1}\left(\int df f P(x,f;t) - mg P(x;t)\right)
\end{equation}
and therefore in the stationary state we need to have $\left<f\right>_{lss}=mg$.

\section{Self-propelled particle in 
a harmonic potential}\label{sec:harm} 

We show in this section that the effective temperature defined through the
long-time diffusive motion of a free self-propelled particle does not
always determine the stationary state probability distribution of the
particle's position in an external harmonic potential. To analyze this
finding a little further, we investigate the particle's position 
auto-correlation function and the linear response to an external perturbation,
and use these analyzes to examine a fluctuation-dissipation relation-based
effective temperature.

\subsection{Stationary state probability distribution}\label{subsec:pd}

For a single self-propelled particle in a harmonic
potential, the equation of motion for the joint probability distribution of the 
position and self-propulsion force has the following form:
\begin{eqnarray}\label{probharm}
\partial_t P(x,f;t) &=& -\frac{1}{\xi_0}
\frac{\partial }{\partial x}\left(\left(f-kx\right)P(x,f;t)\right)
\\ \nonumber && 
+\frac{\partial}{\partial f}\left(\gamma f P(x,f;t) + D_f \frac{\partial}{\partial f}
P(x,v;t)\right),
\end{eqnarray}
where $k$ is the force constant that determines the strength of the potential.

Again, 
we note that the so-called drift coefficients \cite{VanKampen} in Eq. (\ref{probharm})
are linear in $x$ and $f$ and therefore a stationary distribution has a Gaussian form,
\begin{equation}\label{ssprobharm}
P^{ss}(x,f) \propto \exp\left(-ax^2-bf^2-cfx\right),
\end{equation}
where $a=k \xi_0 \left(\gamma + k/\xi_0\right)^2/\left(2 D_f\right)$,
$b= \left(\gamma + k/\xi_0\right)/\left(2 D_f\right)$ and
$c=- k \left(\gamma + k/\xi_0\right)/D_f$.

It follows from Eq. (\ref{ssprobharm}) that the stationary distribution of 
particle's positions is also Gaussian,
\begin{eqnarray}\label{ssprobharmx}
P^{ss}(x) &=& \int df P^{ss}(x,f) 
\\ \nonumber &=&  \left(\frac{a-c^2/(4b)}{\pi}\right)^{1/2}
\exp\left(-\left(a-c^2/(4b)\right)x^2\right),
\end{eqnarray}
where $a-c^2/(4b) = \left(k/2\right)\left(\gamma+k/\xi_0\right)\gamma\xi_0/D_f$.
If we were to define effective temperature through the relation 
$P^{ss}(x) \propto \exp(-V^{ext}(x)/T_{eff})$, we would get 
\begin{equation}\label{Teffharm}
T_{eff}= D_f/\left(\gamma\xi_0\left(\gamma+k/\xi_0\right)\right).
\end{equation}
We note that this effective temperature is different from that defined 
through the long-time diffusive motion of the free self-propelled particle,
Eq. (\ref{Tefffree}). We note, furthermore, that in the present problem
there are two different time scales. First, there is the time scale on
which the self-propulsion force forgets its initial value. As for the 
free particle, this time scale is proportional to $\gamma^{-1}$.
Second, there is the characteristic time scale for the relaxation of
a particle moving in a viscous medium under the influence of a harmonic
potential. This time scale is proportional to $\xi_0/k$. 
If the former time scale is much shorter than the latter,
$\gamma^{-1}\ll \xi_0/k$, the effective temperature (\ref{Teffharm})
coincides with the effective temperature of the free self-propelled particle
(\ref{Tefffree}). In the opposite case, $\gamma^{-1}\gg \xi_0/k$, which 
is the interesting strong self-propulsion limit, the 
effective temperature (\ref{Teffharm})
approaches $D_f/\left(k\gamma\right)$ and can be significantly lower than that of 
the free self-propelled particle (\ref{Tefffree}).

In contrast to the case of a constant external force, the self-propulsion 
distribution of a particle moving under the influence of a harmonic force agrees with
that of the free self-propelled particle,
\begin{eqnarray}
P^{ss}(f) &=& \int dx P^{ss}(x,f) 
\\ \nonumber &=& \left(\frac{b-c^2/(4a)}{\pi}\right)^{1/2}
\exp\left(-\left(b-c^2/(4a)\right)f^2\right)
\end{eqnarray}
where $b-c^2/(4a) = \gamma/(2D_f)$.

However, there is still non-zero local stationary state self-propulsion,
\begin{equation}
\left<f\right>_{lss} = \int df f P^{ss}(f|x) = 
-\frac{cx}{2b} = kx.
\end{equation}
This result is not unexpected since the joint stationary state distribution
(\ref{ssprobharm}) does not factorize into distributions of positions and 
self-propulsions. Physically, this happens because
particles with larger (albeit temporary) self-propulsions are able to venture
farther into the high potential energy regions. 

Finally, we note that, as in the case of a constant external force, 
in the stationary state the current density vanishes,
\begin{equation}
\xi_0^{-1}\left(\int df f P^{ss}(x,f) - kx P^{ss}(x)\right) = 0.
\end{equation}

\subsection{Particle's position auto-correlation function}
\label{subsec:acf}

We use standard methods \cite{VanKampen} to derive coupled equations of motion
for the time-dependent auto-correlation function of the position of the 
self-propelled particle, $\left<x(t)x(0)\right>$, and the
correlation function between the self-propulsion force at time $t$ and
the position at the initial time,  $\left<f(t)x(0)\right>$,
\begin{eqnarray}\label{eomcf1}
\partial_t \left<x(t)x(0)\right> &=& \xi_0^{-1}\left<f(t)x(0)\right> 
- k \xi_0^{-1}\left<x(t)x(0)\right>
\\ \label{eomcf2}
\partial_t \left<f(t)x(0)\right> &=& -\gamma\left<f(t)x(0)\right>
\end{eqnarray} 
Usually, equations of motion for these two functions would involve other,
more complicated time-dependent correlation functions. The equations above
are closed due to the simplicity of the external potential. 
Initial conditions for Eqs. (\ref{eomcf1}-\ref{eomcf2}) are 
$\left<x^2\right>$ and $\left<fx\right>=k\left<x^2\right>$, where,
from Eq. (\ref{ssprobharmx}),  
$\left<x^2\right> = D_f/\left(k\gamma\xi_0\left(\gamma+k/\xi_0\right)\right)$.

Equations of motion (\ref{eomcf1}-\ref{eomcf2}) can be easily solved. The
solution for the particle's position auto-correlation function reads
\begin{eqnarray}\label{acf}
\left<x(t)x(0)\right> =  \left(\frac{\gamma}{\gamma-k/\xi_0}e^{-\frac{k}{\xi_0}t} + 
\frac{k/\xi_0}{k/\xi_0-\gamma}e^{-\gamma t}\right)\left<x^2\right>.
\end{eqnarray}

We note two qualitatively different behaviors in two limiting cases identified
in the previous subsection. If the self-propulsion force relaxation time is
the shortest relevant time scale, $\gamma^{-1}\ll \xi_0/k$,
we get 
\begin{equation}\label{acf1}
\left<x(t)x(0)\right>  \approx 
e^{-\frac{k}{\xi_0}t} \frac{D_f}{k\gamma^2\xi_0}.
\end{equation}
In this case the self-propelled particle's position auto-correlation function
has the same form as the auto-correlation function of a non-active
Brownian particle in equilibrium in an external harmonic potential. 

In the opposite limit, $\gamma^{-1}\gg \xi_0/k$,
we get 
\begin{equation}\label{acf2}
\left<x(t)x(0)\right>  \approx 
e^{-\gamma t} \frac{D_f}{k^2\gamma}.
\end{equation}
We note that in this limit the time dependence
of the particle's position auto-correlation function is slaved to the 
evolution of the self-propulsion force. Interestingly, the auto-correlation function
is independent of the friction coefficient $\xi_0$. 

\subsection{Linear response to an external force}\label{subsec:lr}

To calculate a linear response function we consider the self-propelled particle in 
the harmonic potential and under an influence of a weak time-dependent force. In this 
case, the evolution equation for the joint probability distribution of the position and 
the self-propulsion force has the following form,
\begin{eqnarray}\label{probharmlr1}
\partial_t P(x,f;t) &=& -\frac{1}{\xi_0}
\frac{\partial }{\partial x}\left(\left(f-kx\right)P(x,f;t)\right)
\\ \nonumber && 
+\frac{\partial}{\partial f}\left(\gamma f P(x,f;t) + D_f \frac{\partial}{\partial f}
P(x,v;t)\right)
\\ \nonumber && 
-\frac{1}{\xi_0}\frac{\partial}{\partial x}\left( 
f^{ext}(t) P(x,f;t)\right).
\end{eqnarray}
Here $f^{ext}(t)$ is a weak, time-dependent force which, following the 
analysis of the linear response in equilibrium \cite{RDeL}, we take to be 
position-independent. 

To examine the linear response we linearize Eq. (\ref{probharmlr1}) 
with respect to the external force. In this way we get the following 
equation for the difference between the probability distribution in
the presence of the force and its stationary state form,
$\delta P(x,f;t) = P(x,f;t)- P^{ss}(x,f)$,
\begin{eqnarray}\label{probharmlr2}
\partial_t \delta P(x,f;t) &=& -\frac{1}{\xi_0}
\frac{\partial }{\partial x}\left(\left(f-kx\right)\delta P(x,f;t)\right)
\\ \nonumber && 
+\frac{\partial}{\partial f}\left(\gamma f P(x,f;t) + D_f \frac{\partial}{\partial f}
\delta P(x,v;t)\right)
\\ \nonumber && 
-\frac{1}{\xi_0}\frac{\partial}{\partial x}\left( 
f^{ext}(t) P^{ss}(x,f;t)\right).
\end{eqnarray}
We assume that the force is turned on at $t=0$ and thus the initial condition for 
$\delta P(x,f;t)$ is $\delta P(x,f;t=0)=0$.

Our goal is to calculate the time-dependent change of the particle's position,
$\delta \left<x(t)\right> = \int dx df x \delta P(x,f;t)$. To this end we use
Eq. (\ref{probharmlr2}) to derive coupled equations of motion for
$\delta \left<x(t)\right>$ and $\delta \left<f(t)\right> = \int dx df f \delta P(x,f;t)$,
\begin{eqnarray}\label{eomlr}
\partial_t \delta \left<x(t)\right> &=& \frac{1}{\xi_0}\delta \left<f(t)\right> 
- \frac{k}{\xi_0}\delta \left<x(t)\right> + \frac{1}{\xi_0} f^{ext}(t)
\\
\partial_t \delta\left<f(t)\right> &=& -\gamma\delta\left<f(t)\right>.
\end{eqnarray} 
The initial conditions for these equations are 
$\delta \left<x(t=0)\right> = 0 = \delta \left<f(t=0)\right>$.

Eqs. (\ref{eomlr}) can be easily solved. We get $\delta \left<f(t)\right>\equiv 0$
and 
\begin{equation}\label{xlr}
\delta \left<x(t)\right> = 
\frac{1}{\xi_0}\int_0^t dt' e^{-\frac{k}{\xi_0}(t-t')} f^{ext}(t'),
\end{equation}
and thus the response function is given by 
\begin{equation}\label{lrf}
R(t)=\frac{1}{\xi_0}e^{-\frac{k}{\xi_0}t}.
\end{equation}

\subsection{Fluctuation-dissipation relation}\label{subsec:fdr}

The form of the joint stationary state distribution (\ref{ssprobharm}) 
suggests an effective temperature can be defined for both rapidly evolving
self-propulsion force (in which case $T_{eff}$ is the same as the one defined
through diffusive motion of the free particle) and for the more interesting
slowly evolving self-propulsion force (strong self-propulsion limit).
Physically, in the former case the existence of an effective temperature
is expected but in the latter case it seems to be related to the special form
of the interaction potential. Here, to investigate this a little further, 
we examine a different way to introduce an effective temperature, one that uses 
a fluctuation-dissipation relation (FDR). 

Following a recent review \cite{Cugliandolo} we define a frequency-dependent
fluctuation-dissipation relation-based effective temperature 
\begin{equation}\label{Tefffdrdef}
T_{eff}^{FDR}(\omega) = \frac{\omega \mathrm{Re } C(\omega)}{\chi''(\omega)},
\end{equation}
where $C(\omega)$ is the one-sided Fourier transform of the particle's position
auto-correlation function, 
$C(\omega) = \int_0^\infty e^{i\omega t} \left<x(t)x(0)\right>$,
and $\chi''(\omega)$ is the imaginary part of the one-sided Fourier transform of the 
response function, $\chi''(\omega) = \mathrm{Im }\int_0^\infty e^{i\omega t} R(t)$. 

Using explicit forms of the auto-correlation function and the response function
we get
\begin{equation}\label{Tefffdr}
T_{eff}^{FDR}(\omega) = \frac{D_f}{\xi_0\left(\omega^2+\gamma^2\right)}.
\end{equation}
In principle, the fluctuation-dissipation relation-based 
effective temperature is frequency-dependent and,
thus, the fluctuation-dissipation relation is violated. A more appropriate 
interpretation of Eq. (\ref{Tefffdr}) is that, in the limit of small-frequency
perturbations, $\omega \ll \gamma$, the fluctuation-dissipation relation is recovered
and $T_{eff}^{FDR}(\omega)$ 
coincides with the effective temperature 
obtained from the long-time diffusive motion of the free self-propelled particle.
We note that a similar agreement of effective temperatures measured in different
ways has been found by Loi \textit{et al.} \cite{Loi}. 
We shall emphasize, however, that in the strong self-propulsion limit, 
$\gamma^{-1}\gg \xi_0/k$, the free self-propelled particle-based effective
temperature does not determine the stationary state distribution. 

\section{Discussion}\label{sec:disc}

We shall emphasize three points which we expect to be applicable beyond our
simple toy model. First, the most natural
effective temperature defined through the long-time diffusive motion of 
a free self-propelled particle does not always determine the stationary state 
distribution in an external field, even in the dilute (single particle) limit. 
In other words, the result of Palacci \textit{et al.} \cite{Palacci} is 
highly non-trivial. Second, a well defined
low frequency limit of the fluctuation-dissipation relation-based effective temperature 
may exist and it may be relevant for some properties of the self-propelled
particle \cite{Loi}. However, it does not necessarily determine the 
stationary state of the self-propelled particle in an external potential. 

Third, even though the self-propulsion force evolves 
independently of the state of the self-propelled particle (and independently of the 
interaction of this particle with an external force or with other particles),
non-trivial correlations between self-propulsion force and the particle's 
position can develop. We can also expect that in the case of many interacting 
self-propelled particles, correlations between self-propulsions and 
distances between particles can develop. These correlations imply the appearance of
a non-trivial anisotropic pair distribution function which is thought to
be responsible for the instability of a single phase uniform state in some 
systems of self-propelled particles \cite{Bialke2}. 

\section*{Acknowledgments}
This work was started during a visit to Laboratoire Charles Coulomb of 
Universit\'{e} de Montpellier II. The research in Montpellier was supported by funding
from the European Research Council under the European
Union’s Seventh Framework Programme (FP7/2007-2013) / ERC Grant agreement No 306845.
I thank Ludovic Berthier for many
stimulating discussions and Elijah Flenner for comments on the
manuscript. I gratefully acknowledge the support of NSF Grant No.~CHE 1213401.

\end{document}